\documentclass[%
 reprint,
superscriptaddress,
nofootinbib,
nobibnotes,
 amsmath,amssymb,
 aps,
 prl,
]{revtex4-1}

\usepackage{graphicx}
\usepackage{dcolumn}
\usepackage{bm}
\usepackage{amsmath}
\usepackage{amsfonts}
\usepackage{amssymb}

\begin{document}

\preprint{APS/123-QED}
 
\title{Raman Shifting induced by Cascaded Quadratic Nonlinearities for Terahertz Generation}

\author{Koustuban Ravi}
\email {koustuban@alum.mit.edu}
\affiliation{%
Research Laboratory of Electronics, Massachusetts Institute of Technology
}%
\affiliation{
Center for Free-Electron Laser Science, DESY,Notkestra$\beta$e 85, Hamburg 22607, Germany
}%

\author{Franz X. K\"artner}
\affiliation{%
Research Laboratory of Electronics, Massachusetts Institute of Technology
}%
\affiliation{
Center for Free-Electron Laser Science, DESY,Notkestra$\beta$e 85, Hamburg 22607, Germany
}%
\affiliation{
Department of Physics, University of Hamburg, Hamburg 22761, Germany
}%

\begin{abstract}
We introduce a new regime of cascaded quadratic nonlinearities which result in a continuous red shift of the optical pump, analogous to a Raman shifting process rather than self-phase modulation. This is particularly relevant to terahertz generation, where a continuous red shift of the pump can resolve current issues such as dispersion management and laser-induced damage. We show that in the absence of absorption or dispersion, the presented Raman shifting method will result in optical-to-terahertz energy conversion efficiencies that approach $100\%$ which is not possible with conventional cascaded difference-frequency generation. Furthermore, we present designs of aperiodically poled structures which result in energy conversion efficiencies of $\approx 35\%$ even in the presence of dispersion and absorption.

\end{abstract}

\pacs{Valid PACS appear here}
\maketitle

\section{\label{sec1}Introduction}
Nonlinear optical effects have enabled the generation and conversion of light frequencies from the microwave to the X-ray regions of the electromagnetic spectrum. When one nonlinear optical effect initiates further nonlinear optical effects and their evolution is coupled, they may be collectively referred to as \textit{cascaded} nonlinear optical effects. An inexhaustive list includes \cite {desalvo1992,stegeman1996,ilday2004,moses2006,zhou2014,zhou2017,bache2017,varan1997}.

In the context of cascaded nonlinear optical effects, cascaded second-order processes in quadratic media have garnered significant interest \cite{bache2017}. Of particular interest has been the use of \textit{phase-mismatched} second harmonic generation (SHG) to produce controllable effective third-order nonlinear optical effects on the fundamental wave \cite{desalvo1992,stegeman1996,ilday2004,moses2006,zhou2014,zhou2017}. An extensive set of applications ranging from soliton pulse compression to frequency comb generation and mode-locking \cite{phillips2015,bache2017} have been realized using these \textit{phase-mismatched} second-order processes.

In contrast, another class of cascaded second-order nonlinearities occurs when a large number of highly \textit{phase-matched} second-order processes evolve in concert. This is of great interest to terahertz generation \cite{molter2009,ravi2016,raviCOPA16,hemmer2018,ravi2019} where the repeated frequency down-conversion of an optical or near infrared (NIR) pump photon to multiple terahertz photons via cascaded difference-frequency generation (DFG) was proposed to surmount the single-photon conversion efficiency or Manley-Rowe limit \cite{Golomb04}. The approach has already resulted in the highest optical-to-terahertz energy conversion efficiencies \cite{vicario2014,jolly2019} to date. 

The generation of terahertz radiation is accompanied by a dramatic spectral broadening of the optical pump spectrum. Energy conservation requires the center of mass of the optical spectrum to be red shifted in the case of efficient terahertz generation. However, due to dramatic spectral broadening, both red (Stokes) and blue shifted (Anti Stokes) spectral components are observed in the emergent optical spectrum. Therefore, cascaded DFG in the terahertz range has often been considered to effectively be a self-phase modulation (SPM) effect \cite{gustafson1970,bosshard1995,stegeman1996}.

While cascaded DFG promises to yield optical-to-terahertz conversion efficiencies of several percent \cite{ravi2016,raviCOPA16}, further enhancement of conversion efficiencies are inhibited by the aforementioned spectral broadening. This is due to the following reasons. Firstly, there is an annihilation of terahertz photons via the generation of Anti Stokes spectral components. Secondly, the spectral broadening impedes further terahertz generation due to dispersion \cite{ravi2016}. Thirdly, the spectral broadening causes an increase in the pump intensity which exacerbates laser-induced damage \cite{ravi2019}.

In this letter, we introduce a new regime of cascaded second-order nonlinearties , particularly relevant to narrowband terahertz generation which addresses these issues. We show that in conditions satisfying broadband phase matching, cascaded DFG with an optical pump spectrum comprised of a series of narrowband lines separated by the phase-matched terahertz frequency (or multi-line format) is equivalent to a \textit{Raman shifting} process rather than an SPM process.

In this regime, a continuous red shift accompanied by spectral narrowing of the pump rather than spectral broadening is observed. In the limit of zero absorption and dispersion, this produces optical-to-terahertz energy conversion efficiencies approaching $100\%$. To illustrate the feasibility of the approach under practical conditions, we numerically design an aperiodically poled Lithium Niobate crystal pumped by the aforementioned multi-line format at $1.3\,\mu m$, at a cryogenic temperature of $T=10\,$K, to produce conversion efficiencies of $\approx35\%$.

We begin with an analysis of normalized equations for cascaded DFG of terahertz radiation in Eqs. (\ref{pump_evo}) and (\ref{thz_evo}).
\begin{widetext}
\begin{gather}
\frac{dA_{op}(\omega,z)}{dz} = -j\omega\chi^{(2)}_{eff}\bigg[\int_{0}^{\infty}A_{op}(\omega-\Omega,z)A_{THz}(\Omega,z)e^{j\Delta kz}d\Omega + \int_{0}^{\infty}A_{op}(\omega+\Omega,z)A^{*}_{THz}(\Omega,z)e^{-j\Delta kz}d\Omega\bigg] \label{pump_evo}
\end{gather}

\begin{gather}
\frac{dA_{THz}(\Omega,z)}{dz} =-\frac{\alpha(\Omega)}{2}A_{THz}(\Omega,z)-j\Omega\chi^{(2)}_{eff}\int_{0}^{\infty}A_{op}(\omega+\Omega,z)A_{op}^{*}(\omega,z)e^{-j\Delta kz}d\omega \label{thz_evo}
\end{gather}
\end{widetext}

Equation (\ref{pump_evo}) describes the evolution of the spectral amplitude, $A_{op}(\omega,z)$, of the optical/NIR pump at angular frequency $\omega$ in a quadratic nonlinear medium with an effective second-order susceptibility $\chi^{(2)}_{eff}$. The use of normalized equations make it easier to obtain quantities such as the average frequency shift, as shall be shown subsequently. The first term on the right hand side of Eq. (\ref{pump_evo}) represents the generation of $A_{op}(\omega)$ by DFG between optical and terahertz spectral components, while the second term represents sum-frequency generation (SFG) between optical and terahertz components. The phase-mismatch is given by $\Delta k$. Equation (\ref{thz_evo}) represents the evolution of the terahertz spectral components, $A_{THz}(\Omega,z)$ at angular frequency $\Omega$. The first term in Eq. (\ref{thz_evo}) represents attenuation of the terahertz spectral component at a rate given by the absorption coefficient $\alpha(\Omega)$. The second term in Eq. (\ref{thz_evo}) represents DFG processes between all spectral components of the optical pump. In the absence of absorption, Eqs. (\ref{pump_evo}) and (\ref{thz_evo}) conserve energy.

We will now analytically show that Eqs. (\ref{pump_evo}) and (\ref{thz_evo}) are equivalent to a Raman shifting process. Using a process of multiple-scale analysis \cite{conti2002,moloney1990} or adiabatic elimination \cite{lugiato1984}, Eqs. (\ref{pump_evo}) and (\ref{thz_evo}) can be reduced to a single self-consistent equation for $A_{op}(\omega,z)$ in Eq. (\ref{raman_w}). In particular, we perform the multiple-scale analysis for the case of perfect and \textit{broadband phase matching}, i.e. $\Delta k=0~(\forall\,\omega)$ and finite absorption, i.e. $\alpha>0$.

\begin{widetext}
\begin{gather}
\frac{dA_{op}(\omega,z)}{dz} = -\int_{0}^{\infty}\frac{2\omega_0\Omega{\chi^{(2)}_{eff}}^2}{\alpha(\Omega)}\bigg( A_{op}(\omega-\Omega,z)R(\Omega,z)-A_{op}(\omega+\Omega,z)R^{*}(\Omega,z)\bigg)d\Omega \label{raman_w} 
\end{gather}

\begin{gather}
R(\Omega,z)= \int_{0}^{\infty}A_{op}(\omega'+\Omega,z)A_{op}^{*}(\omega',z)d\omega'\label{P_THZ}
\end{gather}
\end{widetext}

Equation (\ref{raman_w}) is a perturbative solution, valid up to an absorption length $\alpha^{-1}$. In Eq. (\ref{raman_w}), $\omega_0$ is the center frequency of the input pump spectrum. By performing an inverse Fourier transform on Eq. (\ref{raman_w}), it is reduced to the familiar Raman-scattering-like term in the time domain in Eq. (\ref{raman_t}).

\begin{widetext}
\begin{gather}
\frac{dA_{op}(t,z)}{dz} = -\bigg[\chi^{''}(t)\circledast|A_{op}(t,z)|^2\bigg]A_{op}(t,z)\label{raman_t}
\end{gather}
\end{widetext}

In Eq. (\ref{raman_t}), $\chi^{''}(t)$ is used to denote the imaginary part of the Raman gain spectrum and corresponds to the inverse Fourier transform of the first term within the integral of Eq. (\ref{raman_w}).

From Eq. (\ref{raman_w}), we notice that the first term inside the integral on the right-hand side is purely real. This is contrary to the situation for \textit{phase-mismatched} cascaded SHG, where it is purely imaginary. The latter results in spectral broadening and is tantamount to SPM-like effects. However, in the current case of interest, the purely \textit{real and negative} nature of the first term in Eq. (\ref{raman_w}) delineates a transfer of energy from the optical to the terahertz spectrum. Naturally, the rate of this transfer is inversely proportional to the absorption coefficient. In fact, Eq. (\ref{raman_w}) is of the same form as the soliton frequency-shift equation presented by Gordon in \cite{gordon1986}. 

We can further verify this by evaluating the rate of change of the average pump spectrum frequency $\langle\omega\rangle$ using Eq. (\ref{raman_w}).

\begin{gather}
\frac{d\langle\omega\rangle}{dz} = -\frac{8\omega_0{\chi^{(2)}_{eff}}^2}{\alpha}\int_{0}^{\infty}\Omega^2|R(\Omega,z)|^2 d\Omega \label{w_cm}
\end{gather}

Since the integral in Eq. (\ref{w_cm}) is obviously $>0$, it is clear that the $d\langle\omega\rangle/dz<0$, which is proof that there is a continuous red shift.

We evaluate Eq. (\ref{w_cm}) at $z=0$ for the case of a multi-line optical input spectrum containing $N_w$ lines of equal amplitude as shown in Eq. (\ref{equal}).

\begin{gather}
A_{op}(\omega)= \bigg(\frac{\tau}{N_w\sqrt{2\pi}}\bigg)^{1/2}\sum_{m=0}^{N_w-1}  e^{-\frac{(\omega-\omega_0-m\Omega_0)^2\tau^2}{4}} \label{equal}
\end{gather}

In Eq. (\ref{equal}), $\Omega_0$ is the terahertz angular frequency to be generated. The expression for $d\langle\omega\rangle/dz$ is presented below in Eq. (\ref{dw_dz_N}).

\begin{gather}
    \frac{d\langle\omega\rangle}{dz}|_{z=0} = \frac{-8 \pi^{1/2}\omega_0\Omega_0^2{\chi^{(2)}_{eff}}^2}{\alpha\tau}\bigg(\frac{N_w-1}{N_w}\bigg)^2 \label{dw_dz_N}
\end{gather}

From Eq. (\ref{dw_dz_N}), it is evident that the rate of change of $\langle\omega\rangle$ is negative with a magnitude that increases with $N_w$, for a constant pump fluence. A corollary of this is that the average blue-shift is larger for smaller $N_w$. For the case of cascaded optical parametric amplification, $N_w\approx 1$ and therefore $d\langle\omega\rangle/dz|_{z=0} \approx 0$, which is consistent with our previous work \cite{raviCOPA16,hemmer2018}. In fact, simulating different input pump formats using Eq. (\ref{raman_w}) predicts effects for the optical pump which are qualitatively identical to a direct numerical solution of Eqs. (\ref{pump_evo}) and (\ref{thz_evo}) . Now, consider an input pump spectrum of the form given by Eq. (\ref{gauss_amp}). 

\begin{widetext}
\begin{gather}
A_{op}(\omega) = \bigg(\sum_{m=-N}^{N-1}e^{-\frac{2m^2}{N_w^2}}\bigg)^{-1/2}\bigg(\frac{\tau}{\sqrt{2\pi }}\bigg)^{1/2} \sum_{m=-N}^{N-1} e^{-\frac{m^2}{N_w^2}}e^{-\frac{(\omega-\omega_0-m\Omega_0)^2\tau^2}{4}}\label{gauss_amp}
\end{gather}
\end{widetext}

In Eq. (\ref{gauss_amp}), $2N+1$ lines are considered, with an amplitude that decays as $e^{-m^2/N_w^2}$. Using Eq. (\ref{gauss_amp}), we obtain a value for $d\langle\omega\rangle/dz$ given by Eq. (\ref{dw_dz_gauss}).

\begin{gather}
    \frac{d\langle\omega\rangle}{dz}|_{z=0} = \frac{-8 \pi^{1/2}\omega_0\Omega_0^2{\chi^{(2)}_{eff}}^2}{\alpha\tau}\bigg(\frac{\Sigma_{m=-N}^{N-2}e^{-\frac{m^2}{N_w^2}}e^{-\frac{(m+1)^2}{N_w^2}}}{\Sigma_{m=-N}^{N-1}e^{-\frac{2m^2}{N_w^2}}}\bigg)^2 \label{dw_dz_gauss}
\end{gather}

Equation (\ref{dw_dz_gauss}) changes much more dramatically with $N_w$ and is always larger compared to Eq. (\ref{dw_dz_N}). Therefore, it is preferable to use a multi-line format whose amplitudes follow Eq. (\ref{gauss_amp}) rather than Eq. (\ref{equal}).

\begin{figure*}
    \centering
    \includegraphics[scale=0.5]{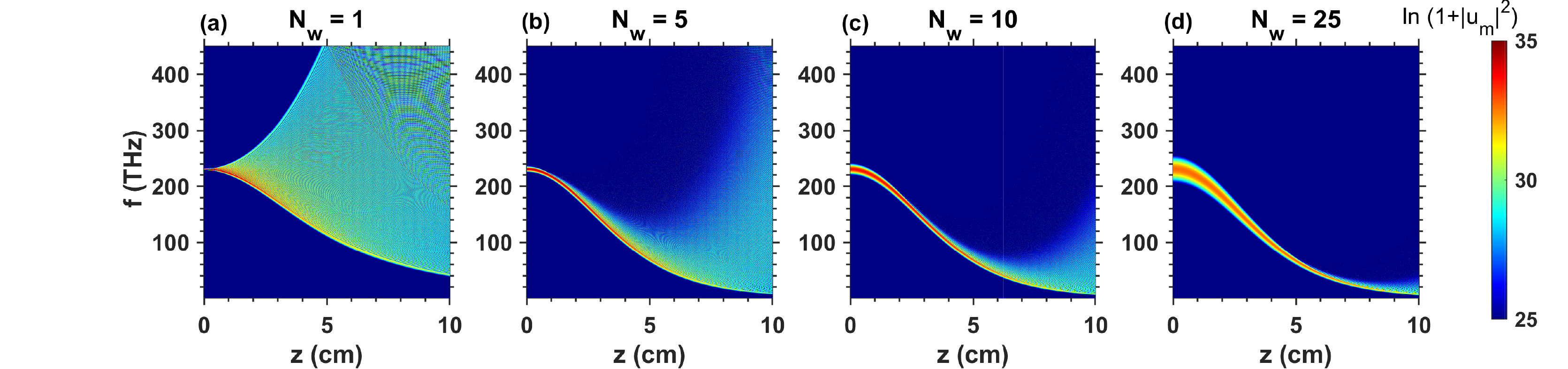}
    \caption{Evolution of multi-line input spectra $u_m(\omega)$ with a Gaussian distribution of amplitudes for various values of $N_w$ in Eq. (\ref{gauss_amp}). (a) For $N_w=1$, significant red and blue shift in frequency is observed. (b) For $N_w=5$, a continuous red shift occurs initially but eventually produces some spectral broadening accompanied by blue shift beyond $z\approx 5\,$cm. (c) For $N_w=10$, continuous red-shift persists for even longer propagation distances. (d) For $N_w=25$, a continuous red shift is maintained over virtually the entire distance of 10\,cm and is visibly accompanied by spectral narrowing rather than spectral broadening.}
    \label{e_ir}
\end{figure*}

Having established the physical concept, we will present simulation results for the case of multi-cycle or narrowband terahertz generation. To alleviate computational cost, we will use a version of Eqs. (\ref{pump_evo}) and (\ref{thz_evo}), which approximate the optical spectrum as a series of discrete lines separate by the generated terahertz angular frequency $\Omega_0$ \cite{raviCOPA16,ravi2019}. These equations have been shown to be in qualitative and quantitative agreement with the exact solutions to Eqs. (\ref{pump_evo}) and (\ref{thz_evo}). In our initial set of simulations, we  exclude dispersion and absorption to show that the Raman-shifting approach will asymptotically approach perfect energy transfer as $N_w$ in Eq. (\ref{gauss_amp}) increases. The evolution of the optical spectra for various $N_w$ is presented in Fig. \ref{e_ir}. A total fluence of $0.35\,$J/cm$^{2}$ is considered and the linewidth of each line in the input optical spectrum is assumed to correspond to a duration of $\tau=200\,$ps.

\begin{figure}
\centering
    \includegraphics[scale=0.3]{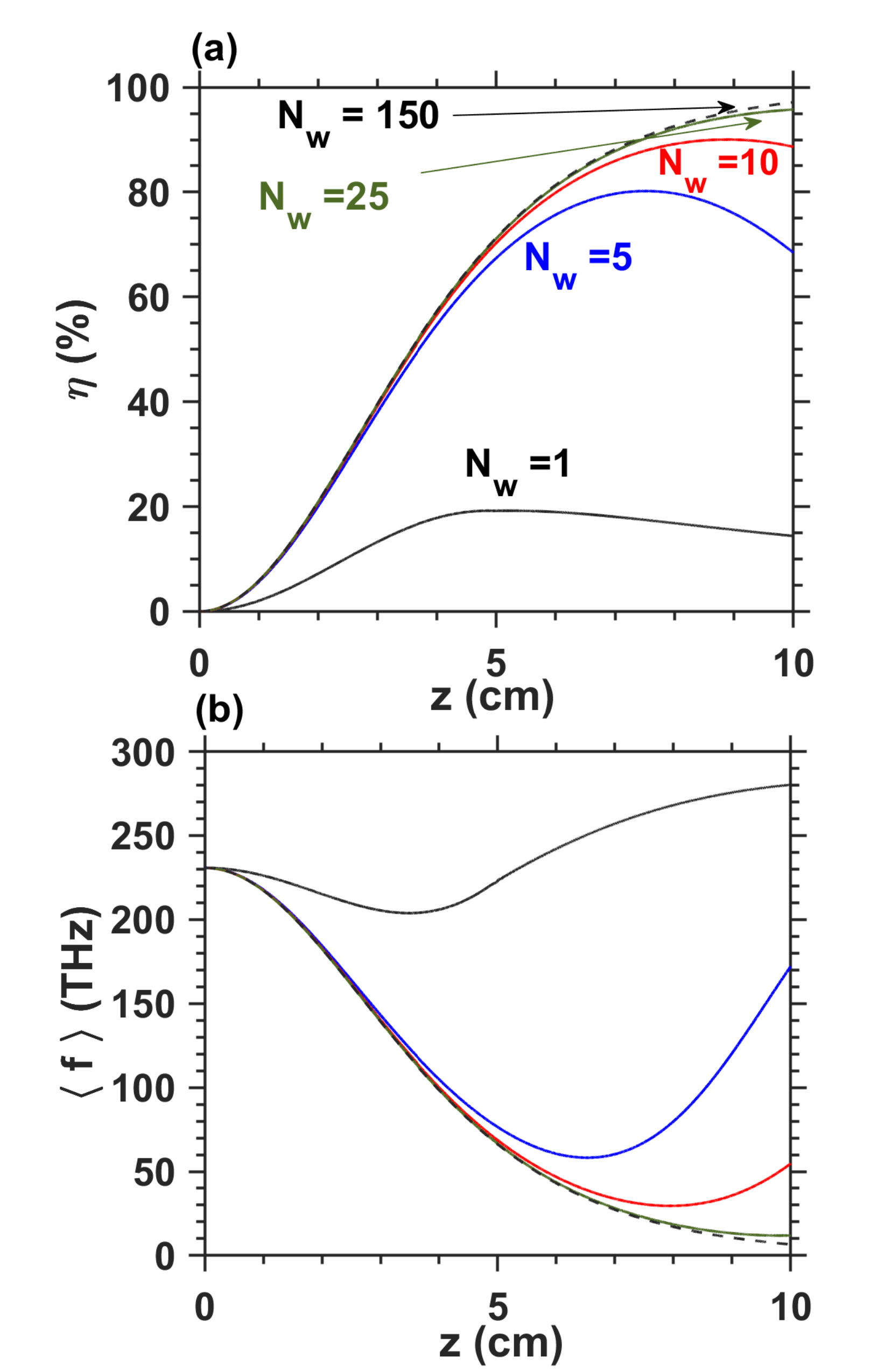}
    \caption{ (a)Optical to terahertz conversion efficiencies and (b) Center of mass of the optical spectrum calculated for various values of $N_w$ in Eq. (\ref{gauss_amp}).}
    \label{eff_wcm}
\end{figure}

 In Fig. \ref{eff_wcm}, we depict the corresponding conversion efficiency, $\eta$ and numerically calculate the evolution of the center frequency of the optical spectrum $\langle\omega\rangle$.
For an input spectrum with $N_w=1$ in Eq. (\ref{gauss_amp}), we notice a significant blue shift in Fig. \ref{e_ir}(a). The simultaneous red and blue shift results in spectral broadening, which is not desired as stated at the outset. In Fig. \ref{eff_wcm}(a), we see that even in the absence of dispersion or absorption, the conversion efficiency reaches only a maximum value of around $20\%$ at $z=5\,$cm and then drops. The case presented in Fig. \ref{e_ir}(a) is the case of conventional \textit{cascaded difference frequency generation}. Correspondingly, in Fig. \ref{eff_wcm}(b), we see that there is an increasing red shift up to $z=5\,$cm. However, this red shift reduces and is accompanied by the reduction in conversion efficiency observed in Fig. \ref{eff_wcm}(a).

In Fig. \ref{e_ir}(b), $N_w=5$. This results in a continuous red shift over a longer distance of $z\approx 7\,$cm before spectral broadening accompanied by blue-shift begins to occur. This is reflected in the conversion efficiency, which rises to nearly $80\%$ in Fig. \ref{eff_wcm}(a) before declining. Since absorption is not included in these calculations, the decline in efficiency merely corresponds to the circulation of energy between the terahertz and optical spectra. Figure \ref{eff_wcm}(b) thus shows a reduction in the net red shift beyond $z=7\,$cm for the case of $N_w=5$.

Figure \ref{e_ir}(c) and (d) depict the evolution of optical spectra for the cases of $N_w=10,25$ respectively. The trend of continuous red shift and lesser spectral broadening and blue shift is observed in these cases. The conversion efficiency in Fig. \ref{eff_wcm}(a) for these cases is correspondingly higher compared to the $N_w=1,5$ cases. For $N_w=10$, a conversion efficiency as large as $90\%$ is attained at $z=8.5\,$cm. However, beyond this distance energy returns to the optical domain as evident by the reduction in red shift in Fig. \ref{eff_wcm}(b). For $N_w=25$, the optical spectrum not only exhibits a continuous red shift but also results in spectral narrowing, which is not seen as distinctly in the previous cases. The spectral narrowing is a consequence of the continuous energy transfer to the terahertz field in the absence of any absorption as evident in Fig. \ref{eff_wcm}(a). Naturally, an incessant red shift is seen for this case in Fig. \ref{eff_wcm}(b). We have also included efficiency and average frequency calculations for $N_w=150$ in Fig, \ref{eff_wcm}. Based on these cases, it is evident that the conversion efficiency asymptotically approaches $100\%$ as $N_w$ increases.

\begin{figure*}
\centering
\includegraphics[scale=0.5]{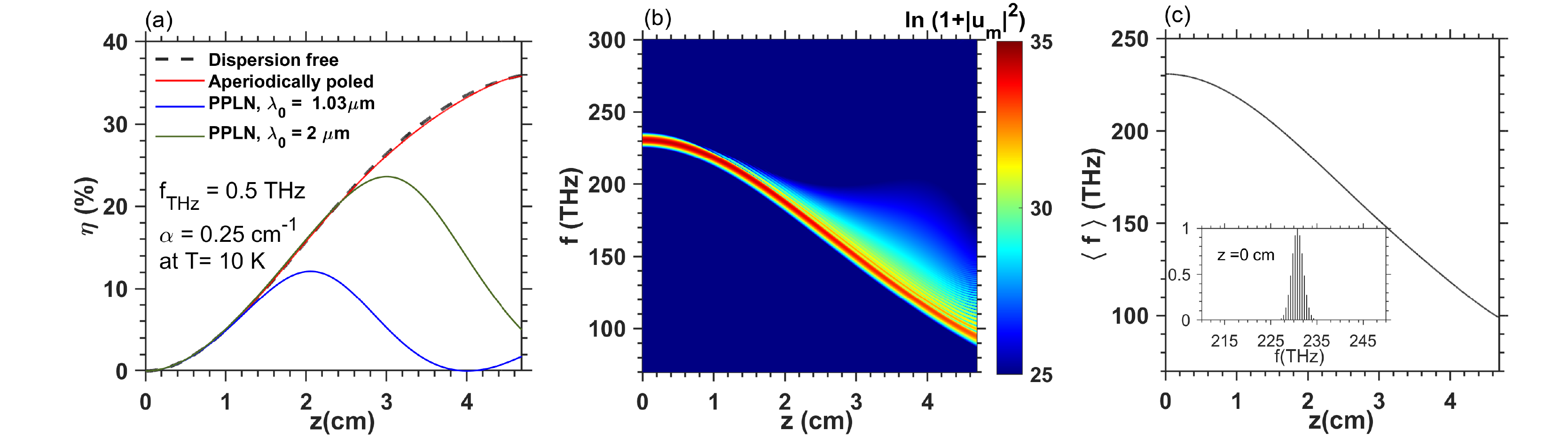}
\caption{Performance in the presence of dispersion and absorption at $T=10\,$K for lithium niobate phase-matched for the generation of 0.5\,THz. (a) For optimized aperiodic poling (red), the conversion efficiency is approximately the same as the dispersion-free case (black,dashed). High conversion efficiencies can also be obtained using periodically poled structures with the pump centered at $\lambda_0=1.03\,\mu$m (blue) and $2\,\mu$m (green). (b) Continuous red-shift in the case of the optimized aperiodic poling, which validates its functionality. (c) The center-of-mass of the optical spectrum corresponding to (b). The input optical spectrum with $N_w=5$ is depicted in the inset.  }
\label{optim_result}
\end{figure*}

Any real system is accompanied by absorption and dispersion and it is therefore imperative that we demonstrate the feasibility of cascaded Raman shifting in such conditions. For illustrative purposes, we assume Magnesium Oxide doped lithium niobate due to their availability with $cm^2$ apertures \cite{ravi2016}. The crystal is assumed to be cooled to $T=10\,$K. For a desired terahertz frequency $\Omega_0/(2\pi)=0.5\,$THz, the absorption coefficient is $\alpha=0.25\,$cm$^{-1}$. Lithium niobate is transparent to optical pump wavelengths between $\lambda_0= 1\,\mu$m and $\approx 3\,\mu$m. Therefore, it is unreasonable to consider any efficiencies which produce red shifting beyond this window. We implemented an optimization algorithm to determine the poling period as a function of distance $z$ for the phase matched generation of $0.5\,$THz. For an initial pump wavelength of $\lambda_0=1.3\,\mu$m, we present the results in Fig. \ref{optim_result}. The choice of center wavelength was based on the ease of convergence of optimization procedures and maximizing conversion efficiency.

 In Fig. \ref{optim_result}(a), we show the conversion efficiency for the case when no dispersion is included (black,dashed line) in the result (absorption is however non-zero). The conversion efficiency for the case of optimized aperiodic poling is depicted in the red curve. As is evident, almost perfect agreement is obtained. The corresponding spectral evolution is presented in Fig. \ref{optim_result}(b) which depicts the desired continuous red shift as evident in the reducing center of mass $\langle \omega\rangle/(2\pi)$ in Fig. \ref{optim_result}(c). The inset in Fig. \ref{optim_result}(c) delineates the input optical spectrum at $z=0$. We have also verified that there is a continuous reduction in the peak intensity of the optical pump in the time-domain, which circumvents laser-induced damage. In addition to the optimized case, conversion efficiency for a periodically poled lithium niobate crystal with central pump wavelengths $\lambda_0=1.03\,\mu$m (blue) and $2\,\mu$m (green) are plotted. The former represents a wavelength for a well developed laser technology while the latter represents a region where the group-velocity dispersion is close to zero for Lithium Niobate. Even in these cases, one obtains high conversion efficiencies of $10\%$ and $25\%$ respectively. The drop in conversion efficiency occurs due to the phase mismatch that sets in as spectral broadening occurs.
 
In conclusion, we introduced a new regime of cascaded quadratic or second-order nonlinearity. We find that under broadband phase matching conditions, cascaded difference frequency generation is akin to a Raman shifting process. In particular, a continuous red shift accompanied by complete energy transfer to the terahertz spectrum can be realized for multi-line input spectra with a large number of lines in the absence of absorption or dispersion. We then theoretically demonstrate a design which can produce a continuous red shift in the presence of absorption and dispersion by an optimized aperiodically poled profile in lithium niobate. Conversion efficiencies as high as $35\%$ can be achieved in this case. Furthermore, since the peak intensity of the optical pulse reduces as terahertz radiation is generated, laser-induced damage can be circumvented. Thus, a new physical mechanism for terahertz generation which can result in very high conversion efficiencies while alleviating limitations due to dispersion and laser-induced damage has been presented.



\bibliography{apssamp}

\end{document}